\documentclass[aps,prb,twocolumn,superscriptaddress,amsmath,amssymb,showpacs]{revtex4-1}
\usepackage{graphicx}
\usepackage{dcolumn}
\usepackage{bm}
\usepackage{color}

\begin{document}

\title{Electronic structure investigation of GdNi using X-ray absorption, magnetic circular dichroism and hard x-ray photoemission spectroscopy}

\author{C.\ W.\ Chuang}
\affiliation{National Synchrotron Radiation Research Center, Hsinchu Science Park, Hsinchu 30076, Taiwan}
\author{H. J.\ Lin}
\affiliation{National Synchrotron Radiation Research Center, Hsinchu Science Park, Hsinchu 30076, Taiwan}
\author{F.~M.~F.~de~Groot}
\affiliation{Inorganic Chemistry and Catalysis, Utrecht University, Universiteitsweg 99, 3584 CG Utrecht, The Netherlands.}
\author{F. H.\ Chang}
\affiliation{National Synchrotron Radiation Research Center, Hsinchu Science Park, Hsinchu 30076, Taiwan}
\author{C.~T.~Chen}
\affiliation{National Synchrotron Radiation Research Center, Hsinchu Science Park, Hsinchu 30076, Taiwan}
\author{Y.\ Y.\ Chin}
\affiliation{Department of Physics, National Chung Cheng University, Chiayi 62102, Taiwan}
\author{Y. F.\ Liao}
\affiliation{National Synchrotron Radiation Research Center, Hsinchu Science Park, Hsinchu 30076, Taiwan}
\author{K. D.\ Tsuei}
\affiliation{National Synchrotron Radiation Research Center, Hsinchu Science Park, Hsinchu 30076, Taiwan}
\author{J.\ Arout Chelvane}
\affiliation{Defence Metallurgical Research Laboratory, Hyderabad 500 058, India}
\author{R.\ Nirmala}
\affiliation{Department of Physics, Indian Institute of Technology Madras, Chennai 600 036, India}
\author{A.\ Chainani}
\affiliation{National Synchrotron Radiation Research Center, Hsinchu Science Park, Hsinchu 30076, Taiwan}

\date{\today}

\begin{abstract}
GdNi is a ferrimagnetic material with a Curie temperature T$_C$ = 69 K which exhibits a large magnetocaloric effect, making it useful for magnetic refrigerator applications. We investigate the electronic structure of GdNi by carrying out x-ray absorption spectroscopy (XAS) and x-ray magnetic circular dichroism (XMCD) at T = 25 K in the ferrimagnetic phase. We analyze the Gd M$_{4,5}$-edge ($3d$ - $4f$) and Ni L$_{2,3}$-edge ($2p$ - $3d$) spectra using atomic multiplet and cluster model calculations, respectively. The atomic multiplet calculation for Gd M$_{4,5}$-edge XAS indicates that Gd is trivalent in GdNi, consistent with localized $4f$ states. On the other hand, a model cluster calculation for Ni L$_{2,3}$-edge XAS shows that Ni is effectively divalent in GdNi and strongly hybridized with nearest neighbour Gd states, resulting in a $d$-electron count of 8.57. 
The Gd M$_{4,5}$-edge XMCD spectrum is consistent with a ground state configuration of S = 7/2 and L=0.
The Ni L$_{2,3}$-edge XMCD results indicate that the antiferromagnetically aligned Ni moments  exhibit a small but finite total magnetic moment ( $m_{tot}$ $\sim$ 0.12 $\mu_B$ ) with the ratio $m_{o}/m_{s}$ $\sim$ 0.11. Valence band hard x-ray photoemission spectroscopy shows Ni $3d$ features at the Fermi level, confirming a partially filled $3d$ band, while the Gd $4f$ states are at high binding energies away from the Fermi level.
The results indicate that the Ni $3d$ band is not fully occupied and contradicts the charge-transfer model for rare-earth based alloys. The obtained electronic parameters indicate that GdNi is a strongly correlated charge transfer metal with the Ni on-site Coulomb energy being much larger than the effective charge-transfer energy between the Ni $3d$ and Gd $4f$ states.

\end{abstract}


\maketitle
\section{Introduction}
Intermetallic alloys consisting of rare-earth and transition metal elements have played a very important role in the development of 
basic and applied solid-state physics and materials science.\cite{Taylor, Wallace, Buschow}  More recently, many intermetallic alloys have shown promising magnetic refrigeration applications related to properties of magnetoresistance, magnetocaloric effect, and magneto-striction.\cite{N1,N2,N3}
The rare-earth $4f$ electrons in intermetallic alloys are usually considered to be localized, while the transition metal $d$ electrons are considered to be delocalized or itinerant. It is well established that in intermetallics, most rare-earth atoms normally exhibit trivalent ground states, except for Ce, Eu, Sm and Yb which can show mixed valency.\cite{Taylor, Wallace, Buschow}  An important question that arises in such intermetallic alloys consisting of rare-earth and transition metal elements is the behavior of element specific magnetism. 

While the rare-earth element is expected to show a local magnetic moment, the behavior of the transition metal magnetic moment is not so clear. It is well-known for many such alloys that when  the concentration of the rare-earth ion increases, the magnetic moments of transition metal ions get reduced.\cite{Taylor} In this picture, it is expected that the transition metal $3d$ band will gradually get filled on increasing rare-earth ion content. This picture is called the charge-transfer model, and it could successfully explain the properties of many, but not all R$_x$M$_y$ alloys (where R = rare-earth and M = Fe, Co, Ni, Cu).\cite{Taylor, Wallace, Buschow} There are some known exceptions to this model such as the alloy system Gd$_{x}$Fe$_{1-x}$ with 0.1 $\leq$ x $\leq$ 0.4.\cite{RTaylor} Also, for GdNi$_2$ with a ferrimagnetic T$_C$ = 85 K, while it was initially concluded that
Ni is non-magnetic in samples prepared at ambient pressure, it was also shown that samples prepared under high pressures of 5 - 8 GPa showed a  reduced T$_C$ $\sim$ 60 K which was attributed to holes in the Ni 3d band indicative of magnetic Ni ions.\cite{Wallace, Slebarski, Tsvyashchenko} These results suggest that the Ni 3d-band is near a filling instability which can be controlled by sample preparation conditions such as high pressure, which results in creating holes in the Ni 3d band leading to Ni magnetic moments.  Subsequently, an XMCD study showed a small but finite spin magnetic moment (m$_{s}$ = 0.14 $\mu_B$) for Ni ions in GdNi$_2$.\cite{Mizumaki}    
In this work, we investigate the electronic structure of GdNi, which has also been debated with respect to the charge-transfer model as discussed in the following. 

GdNi is a binary alloy which undergoes a magnetic transition around T$_C$  = 69 K, and exhibits a large magnetocaloric effect near T$_C$ .\cite{Kumar,Rajivgandhi} The magnetocaloric effect is the thermal response of a material when an external magnetic field is applied or removed under adiabatic conditions. The crystal structure of GdNi is orthorhombic ( CrB type structure, Space group $CmCm$ No. 63 ), and exhibits a trigonal prism arrangement with Gd atoms occupying the prism corners and Ni atoms are positioned at the center of the prism.\cite{B,C,Gignoux} Early studies on the magnetic moments of Ni in GdNi concluded that Ni was non-magnetic, since it was expected that the Ni $3d$ shell can be fully occupied by electrons donated by trivalent Gd ions.\cite{Ursu,Poldy} However, later studies concluded Ni was magnetic, and the Gd moment was ferromagnetically aligned with Ni moments, effectively leading to an excess moment of 7.2 $\mu_B$ compared to the expected theoretical value of a saturation moment of 7 $\mu_B$ for an isolated Gd trivalent ion.\cite{Mallik} Based on band structure calculations, it was shown by Paudyal et al.\cite{Paudyal} that Gd can have a $4f$ derived magnetic moment of 7 $\mu_B$ in GdNi, with Gd 5d electrons also showing a finite moment of 0.3  $\mu_B$, and Ni $3d$ electrons showing a moment -0.1  $\mu_B$. This implied that the Gd moment is antiferromagnetically aligned with Ni moments in GdNi. Paudyal et al. also showed that the anisotropic shifts in lattice constants are responsible for a large spontaneous linear magnetostriction effect of 8000 ppm along the $c$-direction and GdNi can be classified as a giant magneto-striction compound.

From Ni L$_{2,3}$-edge ($2p$-$3d$) XMCD experiments of GdNi carried out earlier,\cite{Yano} it was concluded that the spin magnetic moment of Ni is very small ($m_{s}$ $\sim$ 0.1 $\mu_B$ ) and antiferromagnetically coupled to the magnetic moments of Gd. However, in making the estimate for spin moment from a sum rule analysis, the Ni $3d$ hole density was assumed to be that of GdNi$_2$, based on earlier work.\cite{Mizumaki} The main focus of this work is to quantify the Ni hole density and magnetic moment as well as electronic parameters for describing the electronic structure of GdNi. In this work, we address the hole density of the Ni $3d$ states as well as the magnetic moments associated with the Ni $3d$ and Gd $4f$ states in GdNi using x-ray absorption spectroscopy (XAS) and  x-ray magnetic circular dichroism (XMCD) experiments combined with calculations and a sum rule analysis.\cite{Thole92, Carra93, Chen95,Groot94,Tanaka94}  We confirm trivalency of the Gd ions and consistency of the Gd XMCD with a S = 7/2 and L = 0 ground state. We also carry our bulk sensitive hard x-ray photoemission spectroscopy (HAXPES) of the valence band of GdNi to clarify the character of the occupied density of states near the Fermi level. 
As is well-known,\cite{Fadley,Woicik} HAXPES is a reliable bulk sensitive probe of the intrinsic electronic structure of solids. The HAXPES valence band spectrum of GdNi shows Ni $3d$ features at the Fermi level indicating a partially filled $3d$ band, while the Gd $4f$ states are at high binding energies away from the Fermi level.
The results indicate that GdNi contradicts the charge-transfer model for rare-earth based alloys. The obtained electronic parameters indicate that GdNi is a strongly correlated charge-transfer metal in the Zaanen-Sawatzky-Allen scheme,\cite{ZSA} with the Ni on-site Coulomb energy being much larger than the effective charge-transfer energy between the Ni $3d$ and Gd $4f$ states.

\section{Experiments}

GdNi polycrystalline samples were made by an arc-melting method using high purity constituent elements in argon gas atmosphere. The samples were subsequently annealed at 1173 K for 12 hours in an evacuated quartz tube. The samples were characterized by x-ray powder diffraction and were confirmed to be single phase. Magnetization measurements were carried out using a physical property measurement system ( PPMS, Quantum Design, USA ) and the results confirmed a ferrimagnetic T$_C$  = 69 K, as reported earlier.\cite{Rajivgandhi}  In addition, the magnetization of GdNi tends to saturate above a critical field of about 1.5 Tesla and it does not show a spin-flop behavior.\cite{Kumar, Rajivgandhi} 

XAS and XMCD measurements were carried out at Dragon Beamline ( BL 11A ) of the Taiwan Light Source. The beamline is a bending magnet beamline and the degree of circular polarization $p$ = 0.8. The total electron yield method was used to measure XAS and XMCD across the Gd M$_{4,5}$-edge ($3d$-$4f$) and Ni L$_{2,3}$-edge ($2p$-$3d$) with circularly polarized light and an applied field of +/- 4 Tesla. 
The circularly polarized X-rays were incident normal to the sample surface and the X-ray propagation axis was parallel($\mu_{+}$) or antiparallel($\mu_{-}$) to the applied magnetic field (+/- 4 Tesla). We have used a clockwise circular polarization and a field of +4 Tesla and -4 Tesla to obtain the XAS spectrum as a sum of $\mu_{+}$, $\mu_{-}$ and $\mu_{0}$ ( $\mu_{0}$ corresponds to an applied field perpendicular to the incident beam), and in the first approximation $\mu_{0}$ is taken to be $\mu_{0}$ = [($\mu_{+}$)+($\mu_{-}$)]/2, as is standard procedure.\cite{Thole92, Chen95}  Thus, XAS is proportional to the average of $\mu_{+}$ and $\mu_{-}$ spectra. 

 The photon energy was calibrated using a reference NiO sample and Dy metal sample. The total energy resolution at the Ni $2p$ - $3d$ edge was better than 0.2 eV for the XAS spectra, but for the XMCD spectra, due to the small XMCD signal expected of Ni in GdNi we used a larger slit size to enhance the spectral intensity at the cost of the energy resolution. The reduced energy resolution does not affect the sum rule analysis as the total spectral weight is conserved. For the XAS and XMCD measurements, the sample was polished ex-situ and immediately inserted into the fast entry chamber and pumped down to an ultra-high vacuum(UHV) of better than 8*10$^{-10}$ mbar. The sample was cooled using a liquid He flow type cryostat and the measurements were carried out at T = 25 K. 

We monitored the stability of the Gd XMCD spectra, and the spectra were stable for about 8 hours at T = 25 K. Beyond 8 hours, the Gd XMCD signal got reduced a little. However, we could recover the XMCD signal after heating up the sample to T = 300 K and cooling again to T = 25 K. This indicated that the Gd XMCD signal was sensitive to adsorbed gases but did not change its magnetic and electronic structure and the surface could be refreshed by warming up to T = 300 K. In order to calibrate the energy and confirm absence of oxidation, we simultaneously measured the Ni L-edge of NiO reference sample at T = 300 K in a chamber placed just before the main sample chamber. Similarly,  we also measured O K-edge XAS of NiO reference sample and O K-edge energy range on the GdNi sample. Since the main peak of NiO Ni-L-edge is at 851.8 eV, which is about 1.2 eV below the main peak of Ni L-edge of GdNi at 853.0 eV and since we did not see any feature at 851.8 eV in GdNi sample, we confirmed absence of oxidation on GdNi surface. Further, we could quantify the intensity of O K-edge XAS intensity on sample surface to be less than 3 \% compared to the NiO O K-edge signal. The measured spectra were corrected for the degree of circular polarization and for saturation effects in the TEY mode.\cite{Nakajima}

Valence band HAXPES ( h$\nu$ = 6500 eV ) measurements were carried out at T = 80 K at the Taiwan beamline BL12XU of SPring-8 in Hyogo, Japan. The sample was cooled using a liquid N$_2$ flow-type cryostat.
The overall energy resolution for HAXPES was 0.26 eV as estimated from a fit to the Fermi-edge of gold at T = 80 K, which was also used to calibrate the binding energy scale. For the HAXPES measurements, the sample was cleaved inside the UHV preparation chamber at a pressure of 5*10$^{-9}$ mbar and transferred to the main chamber at 8*10$^{-10}$ mbar for measurements. Although the probing depth of the HAXPES and TEY XAS are $\sim$100 \AA~and $\sim$25 \AA,\cite{Woicik,Nakajima} respectively, since the surfaces were free of oxidation and the XMCD signal compared well with earlier studies on scraped samples,\cite{Yano} the data quality is good and represents the intrinsic electronic structure of GdNi, as is borne out by the results presented in the following. 
 
\section{Results and Discussions}

\begin{figure}[h]
\includegraphics[width=0.5 \textwidth]{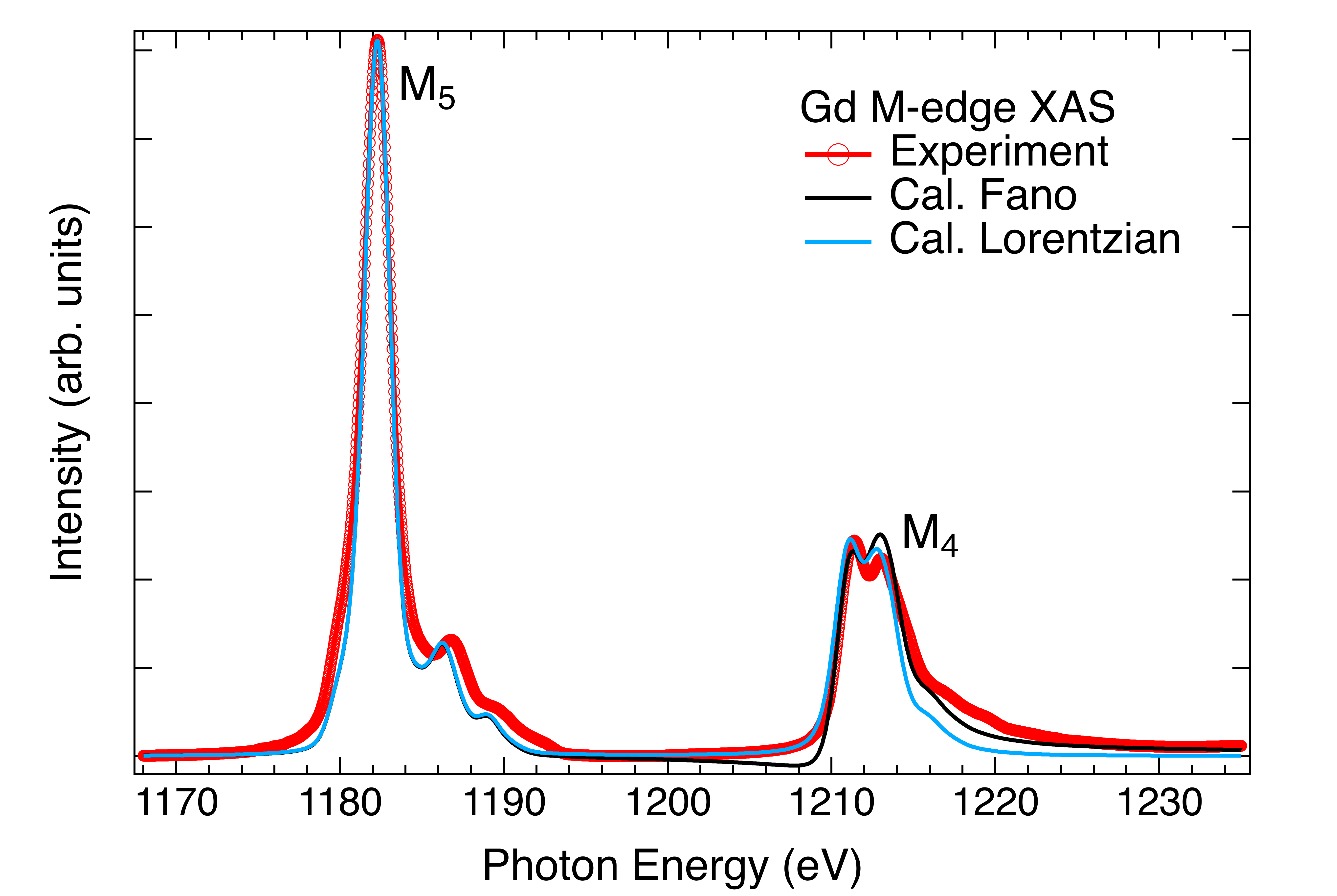}  \caption{ The experimental Gd M$_{4,5}$-edge ($3d$-$4f$) X-ray absorption spectrum of GdNi at T = 25 K compared with atomic multiplet   calculations using two different types of broadenings : a Lorentzian broadening and a Fano broadening.}\label{GdXAS}
\end{figure}

Fig. 1. shows the experimental and calculated Gd M$_{4,5}$-edge ($3d$-$4f$) XAS spectra of GdNi, measured at T = 25 K  in the ferrimagnetic phase of GdNi. 
The experimental XAS spectrum consists of the spin-orbit split M$_5$ and M$_4$ levels separated by about 29 eV. The M$_5$ main peak is observed at an incident photon energy of 1182.3 eV with additional weak features at 1186.8 eV and 1189.6 eV. The M$_4$ features consist of two peaks of nearly equal intensity which are positioned at a photon energy of 1211.4 eV and 1213.0 eV, and a very weak feature is seen at 1217 eV. We used atomic multiplet calculations  using the CTM4XAS program\cite{Groot1} to confirm that Gd is trivalent in GdNi. In order to compare the experimental data with the calculations, we have subtracted out step-functions from the M$_5$ and M$_4$ edges convoluted with a Gaussian
to mimic the transitions from the 3d core state to continuum states. The height of the step functions
were set to a ratio of M$_5$:M$_4$ equal to 6:4, according to the J-value of the core hole state.
Since the excitonic effect on the 4f states is at least 5 eV with respect to delocalised states, 
the step-function was applied from 10 eV above the main peaks.

In Fig. 1, we have plotted two types of  calculated spectra : one with a symmetric Lorentzian broadening and the other with an asymmetric Fano broadening. While all the spectral features are reproduced in both the calculations, there are small differences  between them and also small discrepancies compared to the experimental data. For the Lorentzian and Fano broadened spectra, the weak satellites of the M$_5$ edge are at photon energy positions slightly lower than the energies compared to the experiment. In contrast, for the M$_4$ edge, the satellite energy positions are in fair agreement, but the relative intensities of the multiplet tail at energies above 1215 photon energy are significantly lower in the Lorentzian broadened  spectrum compared to the experiment. This is a general problem known from early work on  M$_{4,5}$-edge XAS spectra of rare-earths and it was shown that a Fano broadening leads to a better match with the M$_4$ edge experimental spectra.\cite{Fano,Thole1985,deGroot1988}  As shown in the Fano broadened spectrum in Fig.1, it is clear that  the multiplet tail intensities above 1215 eV indeed show an improvement compared to the Lorentzian broadened spectrum. While the results confirm the trivalent state of Gd, the finite intensity at high photon energies due to Fano broadening points to a breakdown of the sum rules used for calculating the spin and orbital moments of the Gd ion and are discussed later with the XMCD spectrum in Fig. 4.

\begin{figure}[h]
\includegraphics[width=0.5 \textwidth]{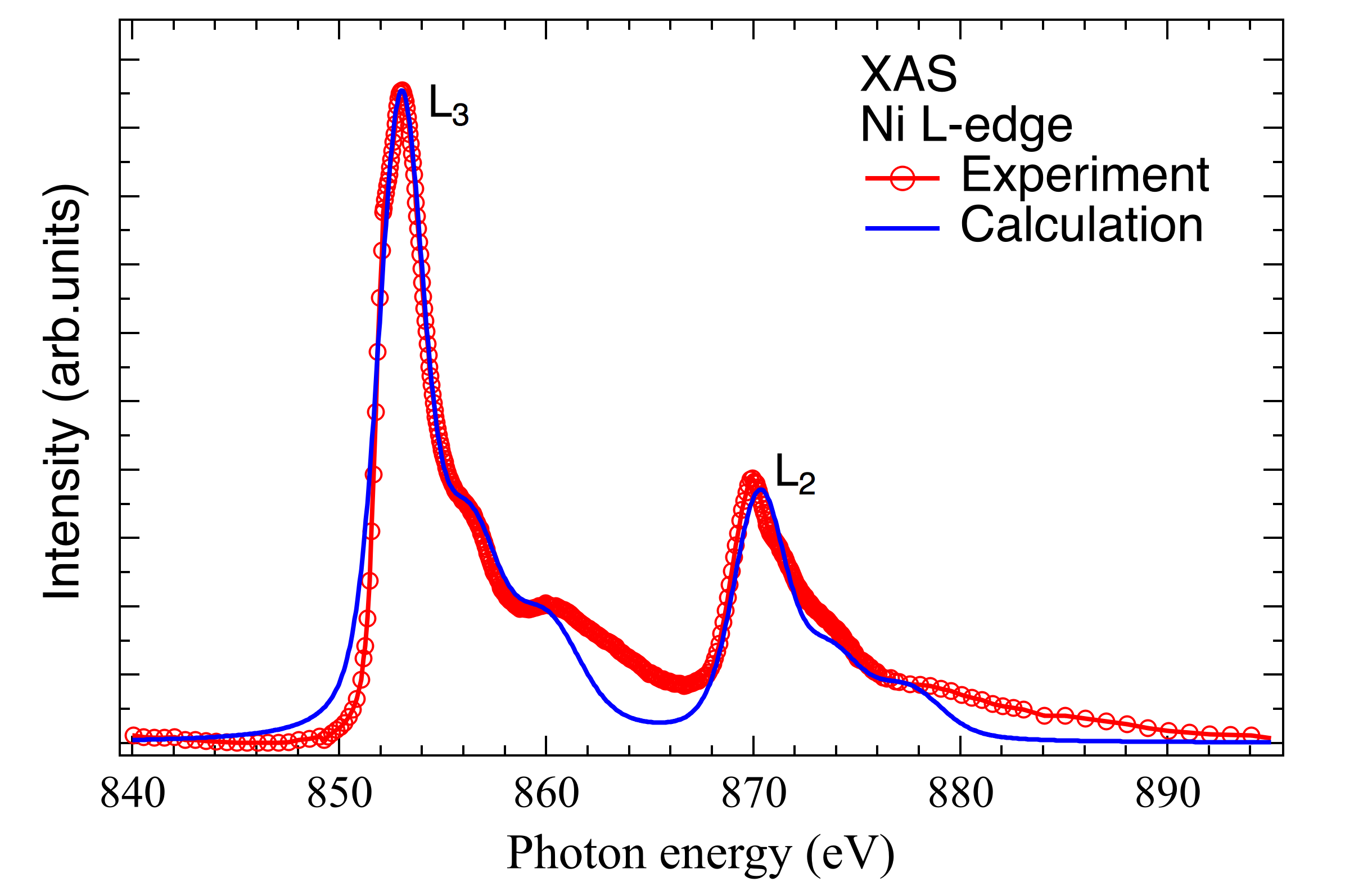}
 \caption{The experimental Ni L$_{2,3}$-edge ($2p$-$3d$) X-ray absorption spectra of GdNi at T = 25 K compared with a calculated spectrum obtained using a metal-ligand cluster model calculation.}\label{NiXAS}
\end{figure}

In Fig.2, we show the experimental and calculated Ni L$_{2,3}$-edge ($2p$-$3d$) XAS spectrum of GdNi measured at 25 K. First of all, it is noted that the spectrum is quite unlike the XAS spectrum of Ni metal. The experimental spectrum consists of spin-orbit split L$_3$ and L$_2$ features separated by an energy of 17 eV. The L$_3$ main peak occurs at an incident photon energy of 853.0 eV and the L$_2$ main peak is observed at an incident photon energy of 869.95 eV. The L$_3$ main peak has satellite features at about 856.05 eV and 860.25 eV, while the L$_2$ main peak has weak shoulders at 873.7 eV and 877.55 eV. We carried out calculations for the Ni $2p$-$3d$ XAS using a NiL$_6$  (L = ligand)  cluster model calculation using the Quanty program.\cite{Haverkort1,Haverkort2,Haverkort3} We used a trigonal prism geometry with D$_{3d}$ symmetry with a formal valency of divalent Ni ion. Although the real local structure of Ni is a regular trigonal prism corresponding to D$_{3h}$ symmetry,\cite{B,C,Gignoux} we needed to use a lower D$_{3d}$ symmetry to get the best match with the experimental data. This implies that the symmetry of the Ni trigonal prism has no mirror symmetry which is present  in the D$_{3h}$ point group. However, our model calculations reproduce all the features of the experimental data and can be considered as a starting point to understand the Ni L$_{2,3}$-edge ($2p$-$3d$) XAS spectra.

The cluster model calculations were carried out by varying the electronic parameters in order to obtain a suitable match to the experimental spectrum. The best match was obtained for the Ni$^{2+}$ configuration with the following electronic parameters:  the charge transfer energy $\Delta$ = -0.6 eV, the on-site Coulomb energy U$_{dd}$ = 6.2 eV, the attractive core hole energy U$_{pd}$ = 10 eV. The value of U$_{dd}$ is typical for Ni containing compounds while U$_{pd}$ is somewhat larger than 
what is normally used, e.g. for NiO, U$_{dd}$ $\sim$6.0 eV while U$_{pd}$$\sim$ 8.0 eV.\cite{Veenendaal}
The relatively large  value of U$_{dd}$ compared to $\Delta$ classifies GdNi as a strongly correlated 
charge-transfer metal in the Zaanen-Sawatzky-Allen (ZSA) scheme which is used to characterize Mott-Hubbard $vis-a-vis$ charge-transfer systems in general.\cite{ZSA} 
The ZSA picture is used to characterize materials in terms of the relative values of the on-site Coulomb energy U$_{dd}$ compared to the charge-transfer energy  $\Delta$ scaled by the hybridization strength, which describes the nature of the lowest energy excitations. When U$_{dd}$ $<$ $\Delta$, the material is called a Mott-Hubbard system with a 
$d-d$ type lowest energy excitation, while if U$_{dd}$ $>$ $\Delta$ it is called a charge-transfer system, where $\Delta$ is the charge-transfer energy determining the lowest energy excitation. The charge transfer energy $\Delta$ is usually defined as the energy separation between the $3d^{n}$ configuration and the charge-transferred 
$3d^{n+1}\underline{L}^{1}$ state, where  $\underline{L}^{1}$ corresponds to one
hole in ligand states.
In our case, the ligand states are the Gd character nearest neighbor states of the Ni site which transfer electrons to the Ni site.
It is noted here that the charge-transfer model of intermetallic alloys is independent of the definition of the charge-transfer metal in the ZSA picture.
The trigonal prism geometry of NiL$_6$ causes a crystal field splitting with three states; a high energy
doublet e$_{g'}$ ($d_{yz}$, $d_{zx}$), an intermediate singlet a$_{1g}$ ($d_{3z^2-r^2}$) and a low energy doublet e$_{g"}$ ($d_{xy}$, $d_{x^2-y^2}$)  states.\cite{Huisman,Burnus}
The a$_{1g}$ and e$_{g"}$ levels can be nearly degenerate or even inverted, depending on the valency as well as the spin-orbit interaction of the central metal ion Ni.\cite{Burnus}
The energy separation between e$_{g'}$ and a$_{1g}$ was taken to be D$_{10}$= 0.45 eV, while the energy separation between a$_{1g}$  and e$_{g"}$ was taken to be D$_{02}$= 1.0 eV
and the hybridization strengths of the $d$ levels with the ligand states were set to V$_{a1g}$ = -2.4 eV, V$_{eg'}$ = 0.65 eV, V$_{eg"}$ = -2.0 eV. A Lorentzian broadening of 0.4 eV full-width at half maximum (FWHM) and an energy dependent Gaussian broadening, with 2.0 eV FWHM for the main L$_3$ and L$_2$ peaks, and 2.5 eV to 3.0 eV for the satellites was used  to calculate the spectrum.
The calculated spectrum matches fairly well with the main peak and satellite structure observed in experiment as shown in Fig. 2. 

We would like to clarify that we first tried many calculations
using the D$_{3h}$ symmetry, as shown in Fig. 3 (Spectra A to J). The best match spectrum obtained with D$_{3d}$ symmetry shown in Fig. 2 is also plotted in Fig. 3 as spectrum K, but with lower broadening parameters to clarify the differences (In Fig. 3, Gaussian broadening is set to 1.5 eV FWHM and Lorentzian broadening is 0.2 eV FWHM for all calculations). In contrast, spectrum A is the best match spectrum we could obtain for  D$_{3h}$ symmetry, but it still fails to reproduce the experimental spectrum. 
In spite of an extensive search with many parameter sets with D$_{3h}$ symmetry, we could not obtain a suitable calculated spectrum to match the experimental
spectrum. In particular, if we got a suitable match for the L$_3$ edge,  the L$_2$ edge did not match (see spectrum A in Fig. 3 ; electronic parameters used : U$_{dd}$= 7.2 eV, U$_{pd}$ = 10.5 eV , $\Delta$ = 0.2 eV, V$_{a1g}$ = -1.732 eV, V$_{eg'}$ = -0.5 eV, V$_{eg''}$ = - 4 eV). In contrast, if the main peaks of the L$_3$ and L$_2$ edges matched the experiment, the intensities of the satellite peaks were too low
 (see spectra E to J in Fig. 3). 
 
 \begin{figure}[h]
\includegraphics[width=0.5 \textwidth]{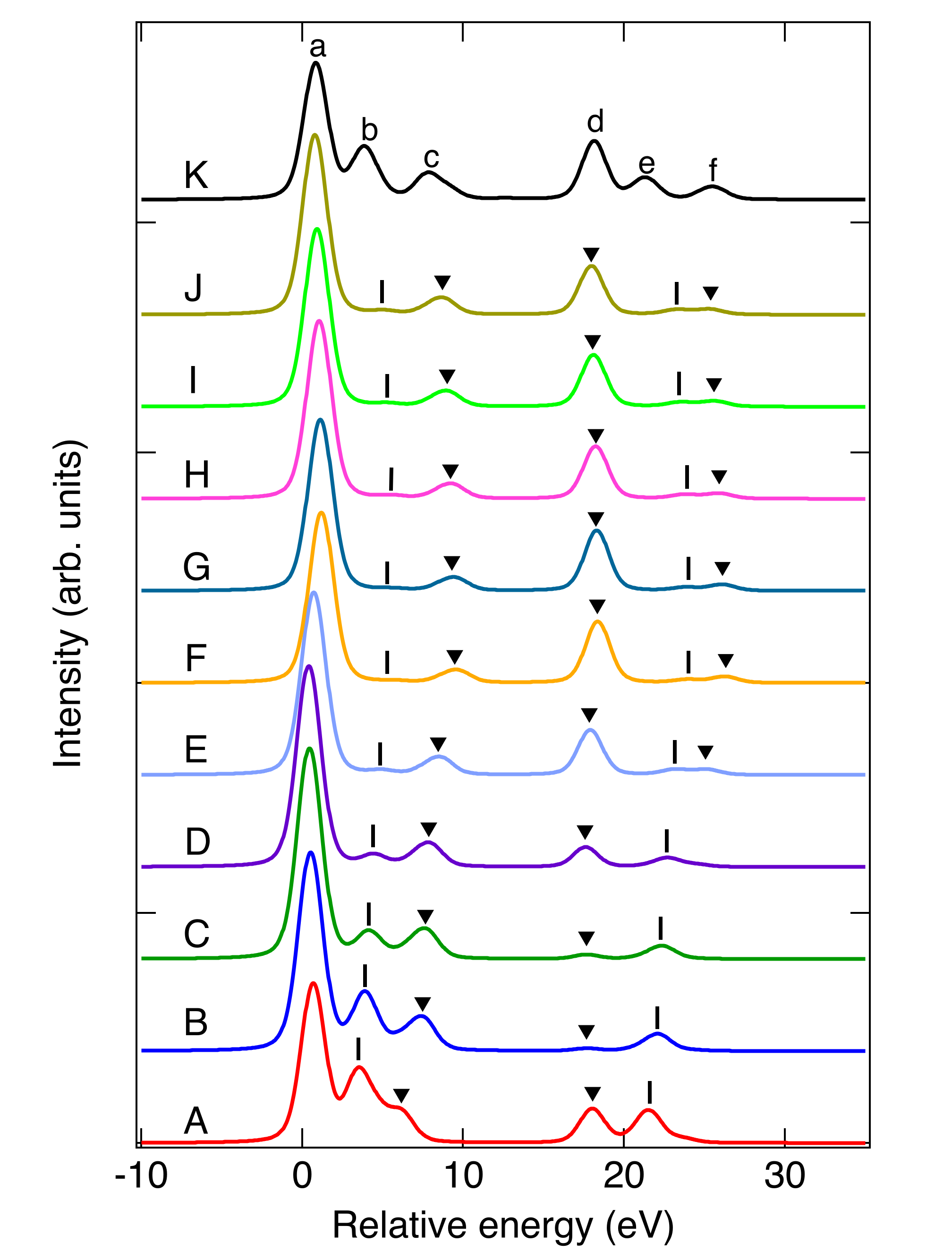}
 \caption{The comparison of Ni L$_{2,3}$-edge XAS spectra obtained using D$_{3h}$ symmetry  (spectra A-J) for different electronic parameters as listed in 
 Table 1 compared with the best match obtained using  D$_{3d}$ symmetry.}\label{Gdlocal}
\end{figure}
 
 In the following, we  briefly describe our attempts to match with the experimental spectrum using D$_{3h}$ symmetry.
Starting with the parameters of spectrum A, we changed U$_{dd}$ and U$_{pd}$ to the same value as used for spectrum K. We obtained spectrum B in which the peaks b, c, d and e got slightly shifted and the intensity of c increased, but the intensity of peak d got reduced. Spectra C and D were obtained by only changing $\Delta$ to -0.2 eV and 0.6 eV, respectively, keeping all other parameters as for spectrum B. The intensity of peak b and f decreased, the intensity of peak d increased, and the positions of peak b, c, and f got slightly shifted. Then, we changed the hybridization energy V$_{a1g}$ to -2 eV and -2.4 eV to obtain spectra E and F. We get a small new feature f, while the intensities of features b, e, and f are very weak. Also, the intensity of peak c decreased, and the intensity of peak d increased. All the peaks were slightly shifted to higher energy. In spectra G, H, only hybridization energy V$_{eg'}$ was changed to 0 eV and 0.65 eV but the spectra hardly changed. Finally, the spectra I, J were obtained by changing hybridization energy V$_{eg"}$ to -3 eV and -2 eV, keeping all the other parameters fixed. The four spectra (G, H, I, J) are very similar. 
Thus, all the parameters of spectrum J (D$_{3h}$) and K (D$_{3d}$) are the same, but clearly spectrum K matches better with the experiment. Hence we felt
using D$_{3d}$ symmetry instead of D$_{3h}$ symmetry, we could obtain a suitable match for energy positions and intensities of the L$_3$ and L$_2$ edges, as well as the satellites seen in the experimental spectrum.
The calculated spectrum K corresponds to a Ni $3d$ electron count of 8.57 in the ground state. This indicates that the Ni $3d$ band is not fully occupied. 
In comparison, it is noted that from a magnetic study of a series of amorphous binary (GdFe, GdCo and GdNi) and ternary alloys of the type
Gd$_{0.2}$(M1$_{x}$M2$_{1-x}$) with M1, M2 = Fe, Co and Ni and x $\sim$0.9 to $\sim$0.2, it was concluded that GdNi corresponds to an effective 
one electron transfer from Gd to Ni.\cite{Taylor1980} This would correspond to the total number of 9 d electrons in GdNi i.e. $3d^{9}4s^{2}$ configuration, starting with  $3d^{8}4s^{2}$ for Ni metal. 
Our results indicate a lower estimate of  8.57 3d-electrons in the ground state in a configuration interaction picture, starting with a formal Ni$^{2+}$ divalent configuration of $3d^{8}$, and admixture from charge transferred states of the type  $3d^{9}\underline{L}^{1}$ and $3d^{10}\underline{L}^{2}$, where $\underline{L}$ corresponds to a hole in the ligand states.
As we show in the following, this results in a small magnetic moment for Ni sites, with slightly different values compared to the results reported by Yano. et al.\cite{Yano} However, in the study of Yano et al., calculations for the XAS Ni $2p$-$3d$ and XMCD spectra were not reported. Thus, our results provide a direct and quantitative measure of the d electron count of Ni in GdNi.

\begin{table}[t]
\caption{Table of electronic parameters for calculated spectra shown in Fig.3.
Error bars for each electronic parameter which individually gives a d-electron count variation of about +/- 0.02, 
and acceptable changes in spectral shape are listed} \label{tb1}
\begin{tabular}{ccccccc}
\hline\\
Parameters : & ~~$\Delta$ & ~~U$_{dd}$ & ~~U$_{pd}$ & ~~V$_{a1g}$ & ~~V$_{eg'}$ & ~~V$_{eg"}$ \\
Error bar&$\pm$0.2&$\pm$1&$\pm$1&$\pm$0.6&$\pm$0.4&$\pm$0.4\\
\hline\\
Spectrum\\
A & 0.2 & 7.2 & 10.5 & -1.732 & -0.5 & -4.0 \\
B & 0.2 & 6.2 & 10.0 & -1.732 & -0.5 & -4.0 \\
C & -0.2 & 6.2 & 10.0 & -1.732 & -0.5 & -4.0 \\
D & -0.6 & 6.2 & 10.0 & -1.732 & -0.5 & -4.0 \\
E & -0.6 & 6.2 & 10.0 & -2.0 & -0.5 & -4.0 \\
F & -0.6 & 6.2 & 10.0 & -2.4 & -0.5 & -4.0 \\
G & -0.6 & 6.2 & 10.0 & -2.4 & 0.0 & -4.0 \\
H & -0.6 & 6.2 & 10.0 & -2.4 & 0.65 & -4.0 \\
I & -0.6 & 6.2 & 10.0 & -2.4 & 0.0 & -3.0 \\
J & -0.6 & 6.2 & 10.0 & -2.4 & 0.65 & -2.0 \\
K & -0.6 & 6.2 & 10.0 & -2.4 & 0.65 & -2.0 \\
\hline
\end{tabular}
\end{table}

\begin{figure}[h]
\includegraphics[width=0.5 \textwidth]{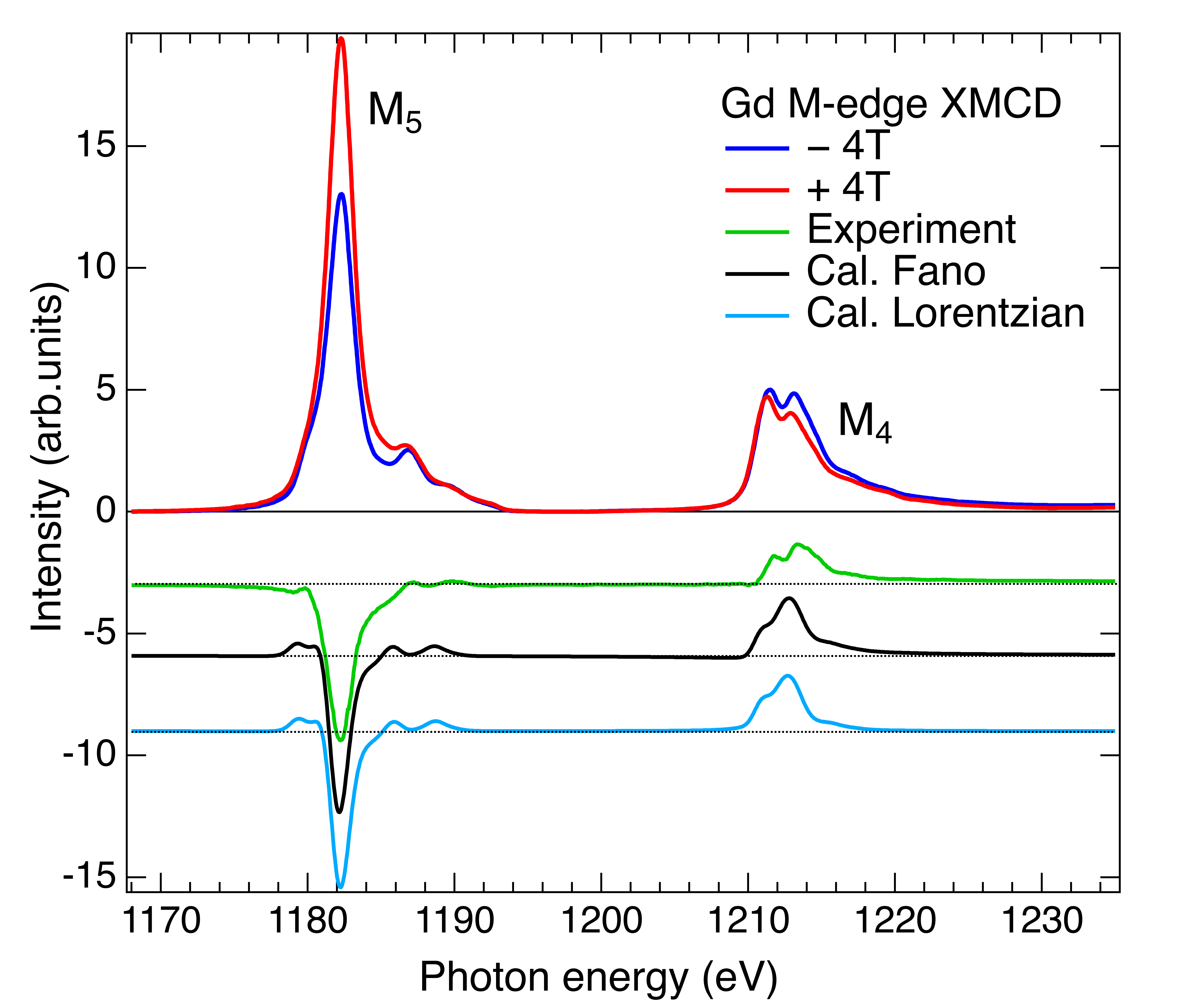}
 \caption{The experimental Gd M$_{4,5}$-edge ($3d$-$4f$) XMCD spectra of GdNi at T = 25 K compared with atomic multiplet   calculations using two different types of broadenings : a Lorentzian broadening and a Fano broadening. The spectra are shifted along the y-axis for clarity.}\label{Gdlocal}
\end{figure}

In Fig. 4, we show the Gd M$_{4,5}$-edge XMCD spectrum measured at T = 25 K, using circularly polarized X-rays and an applied field of +/- 4 Tesla. 
The difference spectrum between the +4 Tesla and -4 Tesla spectra shows a clear XMCD signal as seen in Fig. 4. 
The spectral features are very similar to earlier XMCD spectra of GdNi,\cite{Yano} as well as other Gd containing materials like GdNi$_2$,\cite{Mizumaki} Ce$_{0.5}$Gd$_{0.5}$Ni,\cite{Okane} the metallacrown materials GdNi$_8$,\cite{A}  etc.
The similarity of the XMCD spectra in different materials indicates consistency with a localized picture for the Gd$^{3+}$ ions. We have used the CTM4XAS program with the same electronic parameters as for the calculated XAS spectrum shown in Fig. 1 to also calculate the XMCD spectrum. Here again, we have calculated the 
XMCD spectra with a symmetric Lorentzian broadening and an asymmetric Fano broadening.
The calculated XMCD spectra for both cases are very similar indicating that although the XAS spectrum was better explained by the Fano broadened spectrum, the XMCD spectrum is not sensitive to the type of broadening used. The calculated XMCD spectra correspond to a ground state configuration of S = 7/2 (S$_z$ = -3.465 $\mu_B$ ) and L = 0 (L$_z$ = -0.0345 $\mu_B$).  L$_z$ is not exactly zero due to a combination of $4f4f$ multiplet interactions and the $4f$ spin-orbit coupling.
However, if we use a Fano broadening to simulate the XAS and XMCD spectra, this leads to a small dip before the M$_{4}$ edge and finite intensity at photon energies far above the M$_4$ edge in XAS and XMCD spectra. This means that we cannot set the XMCD intensity to zero before and after the edge. Further, the integral of the XAS and XMCD spectra do not saturate at high photon energies. In turn, this implies that the orbital magnetic moment, m$_{o}$ cannot be 0 even for Gd$^{3+}$ ions with L = 0, leading to an effective breakdown of the sum rule analysis.\cite{Thole92,Carra93,Chen95}
It is interesting to test this proposition quantitatively on other rare-earth systems and future work in this direction  would be valuable. 
We have nonetheless carried out a sum rule analysis using the atomic multiplet calculation results and compared it with the experimental XAS and XMCD data. 
We first normalized the calculated  XAS spectrum with the experimental spectrum. Using the same normalization factor for the theoretical XMCD intensity results in a higher intensity of the calculated spectrum compared to the experimental XMCD spectrum. The atomic calculation results need to be reduced to 92.6\% in order to match the experimental XMCD spectrum. Based on this, the experimental estimate of an effective spin magnetic moment m$_{s,eff}$ = 2S$_{z,eff}$ = 6.42 $\pm$0.1$\mu_{B}$, i.e. it is reduced compared to the 7 $\mu_{B}$ corresponding to the S = 7/2 ground state. Similarly, the effective orbital magnetic moment was reduced to m$_{o,eff}$ = 0.032 $\pm$0.01$\mu_{B}$. We attribute the reduced effective spin moment m$_{s,eff}$ to the role of crystal field effects, finite temperature of the measurement as well as the fact that we needed a Fano type broadening to better explain our experimental data for the multiplet tail intensities above 1215 eV (Fig. 1). It is noted that in a recent study on in-situ deposited amorphous thin films of Co$_x$Gd$_{1-x}$, a reduction of Gd spin moment with m$_{s}$ = 3.8 $\mu_{B}$ was reported.\cite{Bergeard} 

\begin{figure}[h]
\includegraphics[width=0.5 \textwidth]{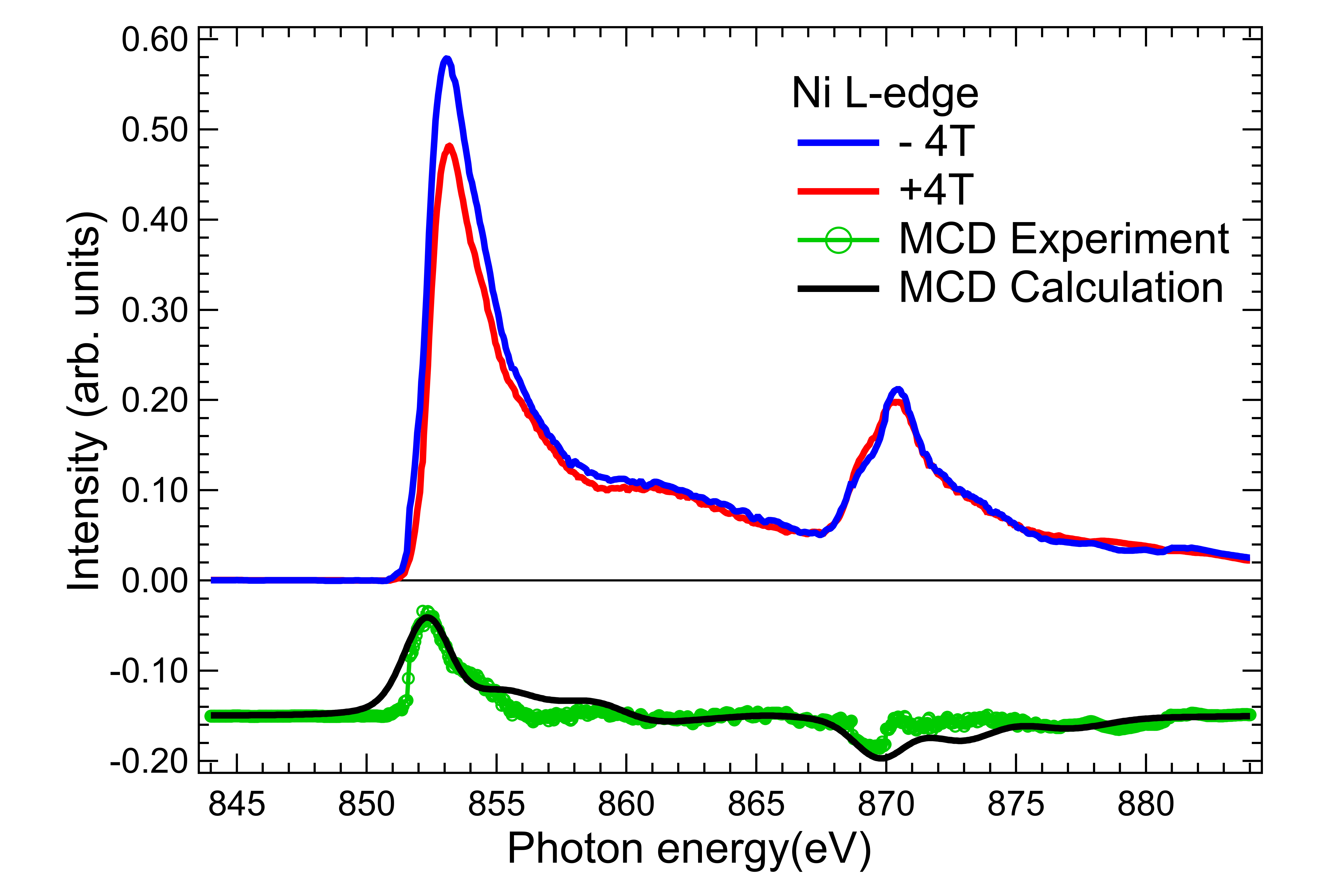}
 \caption{The experimental Ni L$_{2,3}$-edge ($2p$-$3d$) X-ray magnetic circular dichroism spectra of GdNi at T = 25 K compared with calculated spectrum obtained using a metal-ligand cluster model calculation. The spectra are shifted along the y-axis for clarity}\label{NiXMCD}
\end{figure}

Fig. 5 shows the Ni XMCD spectra measured at T = 25 K, using circularly polarized X-rays and an applied field of +/- 4 Tesla. The data shows a clear XMCD signal as seen in the +/- 4 Tesla difference spectrum.  
The sign of the Ni L$_{2,3}$-edge ($2p$-$3d$) XMCD results indicate that the Ni moments are antiferromagnetically aligned with the Gd moments.
The Ni XMCD data are slightly different from that of Yano et al. for the main L$_3$ peak which has a higher energy weak shoulder in our case but was a single peak in the work of Yano et al.\cite{Yano}. However our Ni XMCD data matches well with the Ni XMCD data of fractured single crystal Ce$_{0.5}$Gd$_{0.5}$Ni reported by the same group later.\cite{Okane} We have used the same electronic parameters as for the calculated XAS spectrum to also calculate the XMCD spectrum. As seen in Fig. 5, the calculated spectrum reproduces all the features of the XMCD experimental data, although the intensities for the weak features don't match well with the calculations. We have estimated the  spin and orbital magnetic moments  m$_{s}$  and m$_{o}$ for Ni ions using a sum rule analysis\cite{Thole92,Carra93,Chen95} and we obtained m$_{s}$ = 0.11 $\pm$0.01$\mu_{B}$ , m$_{o}$ = 0.012 $\pm$0.01$\mu_{B}$, giving  $m_{tot}$ $\sim$ 0.12 $\pm$0.01$\mu_B$. These values are slightly different than previously estimated values of m$_{s}$ = 0.09 $\mu_{B}$ and m$_{o}$ = 0.014 $\mu_{B}$.\cite{Yano} It is also noted that Yano et al. reported a ratio m$_{o}$/m$_{s}$ = 0.16,\cite{Yano} and we obtain a ratio of m$_{o}$/m$_{s}$ = 0.11. 

\begin{figure}[h]
\includegraphics[width=0.5 \textwidth]{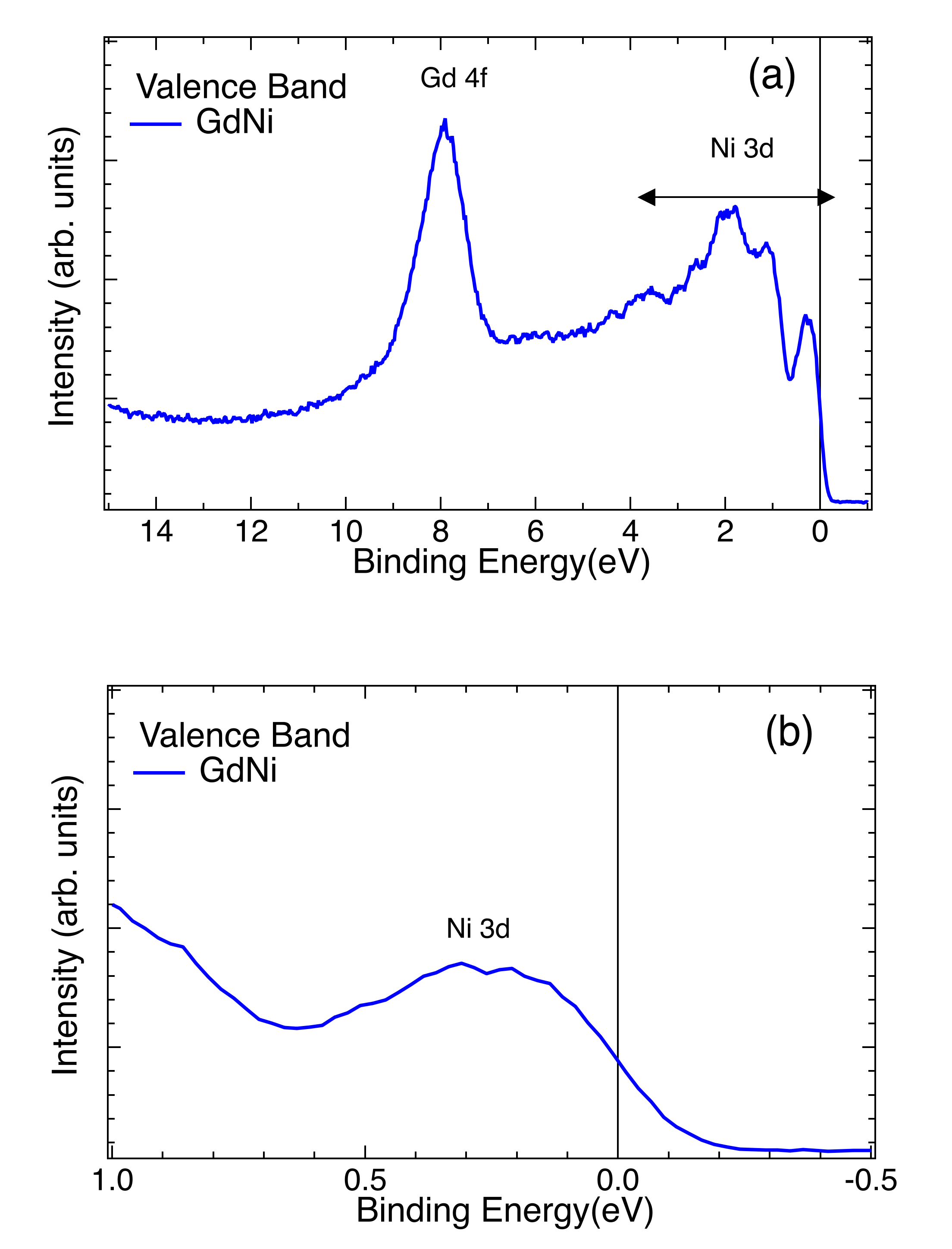}
 \caption{(a) The experimental hard x-ray wide valence band photoemission spectrum of GdNi at T = 80 K shows the Gd $4f$ states at high binding energies and the Ni $3d$ dominated states at the Fermi level and within 4 eV binding energy. (b) The near Fermi level region plotted on an expanded binding energy scale shows the Fermi level lies at the mid-point of the leading edge of the Ni 3d feature.}\label{VB}
\end{figure}

Finally, in Fig. 6(a) and (b), we plot the wide valence band and the near Fermi level HAXPES spectrum of GdNi measured at T = 80 K, respectively.
The  HAXPES valence band of GdNi shows the Gd $4f$ states as a broad intense peak centered at a high binding energy of 8 eV. Further, the Ni $3d$ dominated states occur at the Fermi level and within 4 eV binding energy : a sharp peak is seen at 0.25 eV binding energy ( the near Fermi level spectrum is drawn on an expanded binding energy scale in Fig. 6(b)), followed by higher intensity features at 1.1 eV and 2.0 eV, and weak features at 2.7 eV and 3.5 eV binding energy. This assignment is qualitatively consistent with
electronic structure calculations of the Ni 3d density of states obtained using a tight-binding linear linear muffin-tin orbitals (TB-LMTO) method,\cite{Rusz} although  the sharp peak at the Fermi level observed in our experimental data occurs at about 1.3 eV binding energy in the calculations. The calculations further indicated that the Gd 4f occupied states occur at about 4.8 eV binding energy and the weaker intensity Gd 5d states are spread between the Fermi level and 5 eV binding energy. Hence it is possible that the weak features at 2.7 eV and 3.5 eV binding energies may also have contributions from Gd 5d states which overlap the Ni 3d states. On the other hand,
soft x-ray photoemission studies on several Gd based intermetallics\cite{Kowalczyk,Pop,Kwiecen,Bajorek, Coldea}  containing Gd and Ni, such as GdNi$_{5-x}$Al$_{x}$, Gd$_{3}$Ni$_{8}$Al, GdNi$_{4}$B, GdNi$_{4}$Si, Gd(Ni$_{1-x}$Co$_{x}$)$_{3}$  and Gd(Ni$_{1-x}$Fe$_{x}$)$_{3}$ have shown a broad Gd $4f$ derived feature at 8 eV binding energy, while the Ni $3d$ states occur between Fermi level and about 4 eV binding energy. Our results indicate strong similarity of the spectral features of Gd $4f$ and Ni $3d$ states as observed in the series  Gd(Ni$_{5-x}$Al$_{x}$), Gd(Ni$_{1-x}$Co$_{x}$)$_{3}$  and Gd(Ni$_{1-x}$Fe$_{x}$)$_{3}$. However, one important difference in GdNi compared to Gd$_{3}$Ni$_{8}$Al, GdNi$_{4}$B and GdNi$_{4}$Si is that, we see a sharp lowest binding energy Ni $3d$ derived feature at 0.25 eV cutting the Fermi level,(Fig. 6(b)) while the lowest binding energy Ni $3d$ feature in Gd$_{3}$Ni$_{8}$Al,
GdNi$_{4}$B and GdNi$_{4}$Si is at about 1 eV binding energy and shows weak intensity at the Fermi level.\cite{Pop,Kowalczyk} Our results thus indicate that the Ni $3d$ band occurs at the Fermi level and is partially occupied. While the Ni $3d$ states would participate in charge and thermal transport processes, the very small magnetic moment of Ni in GdNi suggests it would result in a weak contribution to the magnetocaloric effect, while the large Gd spin moments would play a dominant role in the magnetocaloric properties of GdNi.

\section{Conclusion}
In conclusion, we have carried out XAS and XMCD of  GdNi in the ferrimagnetic phase at T = 25 K. The Gd M$_{4,5}$-edge and Ni L$_{2,3}$-edge XAS  and XMCD spectra could be analyzed using atomic multiplet and cluster model calculations, respectively. The Ni L$_{2,3}$-edge XAS and XMCD experimental spectra are fairly consistent with the cluster model calculations carried  out for the D$3d$ local symmetry. We could quantify the hole density of the Ni $3d$ states as well as the magnetic moments associated with the Ni $3d$ states in GdNi. We also confirmed the trivalency of Gd ions and consistency of the Gd XMCD with a S = 7/2 and L = 0 ground state. The HAXPES valence band spectrum shows a Ni $3d$ character sharp peak at the Fermi level, consistent with a partially filled $3d$ band, while the Gd $4f$ states are at high binding energies away from the Fermi level.
The results indicate that the Ni $3d$ band is not fully occupied and contradicts the charge-transfer model for rare-earth based alloys. 

\section{Acknowledgement}
HJL thanks the Ministry of Science and Technology of the Republic of China, Taiwan, for financially supporting this research under Contract No.MOST 106-2112-M-213 -003 -MY3. AC thanks the Ministry of Science and Technology of the Republic of China, Taiwan, for financially supporting this research under Contract No. MOST 106-2112-M-213-001-MY2.

\end{document}